# Construction of hazard maps of Hantavirus contagion using Remote Sensing, logistic regression and Artificial Neural Networks: case Araucanía Region, Chile

Gabriel A. Álvarez A.[1], Luis A. Fernández[2], Renato A. Salinas S.[3]
Author affiliations: University of Antofagasta, Antofagasta, Chile (G. Álvarez, and L. Fernández); University Santiago de Chile (R. Salinas).
These authors contributed equally to this article.

## Abstract

In this research, methods and computational results based on statistical analysis and mathematical modelling, data collection in situ in order to make a hazard map of Hanta Virus infection in the region of Araucania, Chile are presented. The development of this work involves several elements such as Landsat satellite images, biological information regarding seropositivity of Hanta Virus and information concerning positive cases of infection detected in the region. All this information has been processed to find a function that models the danger of contagion in the region, through logistic regression analysis and Artificial Neural Networks.

## 1. Introduction

Considering the study of vector-borne diseases, it is common today to consider the relationship of the environment with the development of the disease, which is termed as eco-epidemiology (*1*). If, in addition, its geographical position is incorporated, its physical and / or biological variations, and mapping limitation then it becomes what is known as panoramic epidemiology (*2, 3*).

In recent times, there has been considerable input from technologies such as Geographic Information Systems (GIS) and Remote Sensing (RS) to environmental analysis and to the study of infectious diseases, thereby configuring satellite epidemiology (1).

As an example of these contributions we can mention the use of Landsat images and GIS for the study of the relationship between mice bearing Nameless virus (SNV) and its ecological environment (*4*); SPOT imagery and DEM Shuttle Radar Topography Mission (SRTM) in the case of rodents carrying Calabazo virus (*5*); the use of satellite technology for models predicting cholera outbreaks, and the study of rodents carrying hantavirus pulmonary syndrome (*6*). MODIS Images for the study of the density of rodents carrying SNV virus (*7*), Geographical Information Systems to examine the potential distribution of hantavirus pulmonary syndrome caused by the Andes virus (ANDV) in southern Argentina (*8*), Geographical Information Systems for the analysis of the characteristics space - time of haemorrhagic fever with renal syndrome (HFRS) in Hubei province, China (*9*), among others.

## 2. General aspects of Hanta Virus

The first reports of Hanta Virus dates back to 1951 during the Korean War (*10*), while in America, the first cases were recognized in 1993 in the south of the United States (*11*).
In this regard, two syndromes have been described clinically which are caused by Hantavirus whose reservoir of different varieties are wild rodents (*12*), haemorrhagic fever with renal syndrome (HFRS) endemic to Asia and Europe, and Hantavirus Pulmonary Syndrome (SPH) that only occurs in America, with cases detected in Argentina, Bolivia, Brazil, Canada, Chile, Mexico, Panama, Paraguay and Uruguay (*13*).

In Chile, the first confirmed case of SPH was diagnosed in October 1995 in the town of Cochamó province of Llanquihue, X Region (*13, 14*), the Andes virus being the etiologic agent, and whose primary reservoir is the Oligoryzomys longicaudatus or long-tailed mouse (*13, 14*), which is present from the Atacama region to the Aysen region *(14)* plus four others species being found seropositive to the virus (*16*).

As their habitat mainly found in natural environment, with abundant forest, dense vegetation in damp places near water sources such as rivers or lakes (*12, 14*), with occurrences up to 2000 meters above sea level.

The mode of transmission is mainly by inhaling particulate matter associated with excretions and droppings of rodents or by direct contact or by ingestion of contaminated food or water (*17*).

## 3. Hypothesis
Considering the general aspects, the region of Araucania and southern Chile it is clear a reservoir of Hanta Virus, probably supported by specific environmental conditions, the presence of vegetation, humidity and limited to a specific geographical altitude. External factors can be measured remotely using satellite images and others factors directly with field measurements. If there is a relationship between environmental factors with positive infections in the region of Araucania, then we can generate two models: one, with logistics regression and other using neural networks, to generate two maps of danger of contagion.

## 4. Materials
The work of gathering information comes from several sources. The satellite images were provided by the Center for Space Studies at the University of Chile (EEC) for the Landsat satellite platform sensor 5 TM, of 15 January 1987 and 29 November 2001. The statistics of positive cases of Hanta virus contagions in the region of Araucania, the ministerial regional secretary of the Araucania region. Rodents statistics provided by the Chilean Ministry of Health (*17*). NASA DEM was obtained for generating of heights.

## 5. Methodology
A methodology to construct hazard maps of spread of Hanta Virus has been partially described in the literature for events of hanta virus in Chile. In Chile, infections occur regularly and the Ministry of Health has resources for scientific research, however, in general, these resources have been focused on research in the area of infection and to quantify areas and vector density of contagion. The transformation of the information taken in the field via some statistical process and incorporated into a layer is particularly important and requires the advice and knowledge of a specialist in order to establish a real equivalence between data and established probability.

## 5.1 Construction of layers
This process involves the creation of information for model development, building layer or layers of interest being necessary from the information gathered, for it states:

**Layer HANTA**. In general, it is especially difficult to secure a place of transmission of hanta virus since the early stages of infection are asymptomatic, so we only have general references as the name of the locality or place most likely where the victim of the contagion was in the days before the onset of the first symptoms. Therefore, for the geo-referencing of information provided by the Seremia of the Araucania region it was necessary to use cartographic information such as letters of the Military Geographical Institute (IGM), satellite information from Google Earth and site visits for geographic spatial determination by GPS mapping. Considering then that both vegetation distribution and geographical features are similar in predictable ranges established that the geomorphology in the near 500 m radius. It is a homogeneous distribution of infectious vector. The information was generated using Visual Basic © and the result was a set of 46,274 pixels; of these 23,120 were positive cases and a number of pixels corresponding to 23,154 negative cases homogeneously distributed in the numeric array representing the cell training.

**Layer NDVI.** Layer realizes the presence of vegetation and which is obtained from the ratio of the subtraction and addition of the energy at wavelengths corresponding to red and near infrared data, using satellite images. This information was obtained directly from Envi © algorithms.

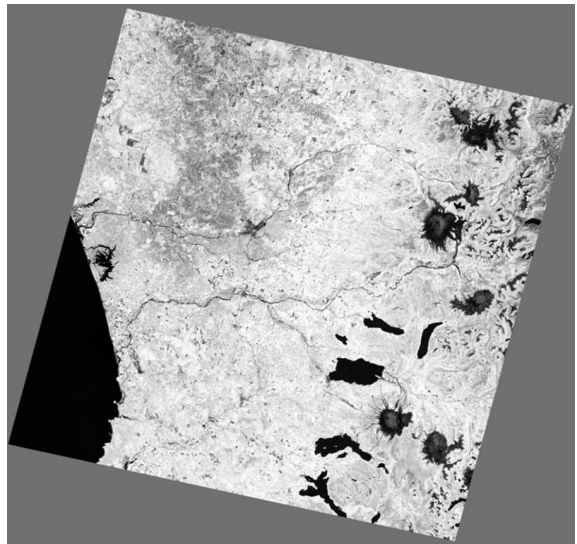

Figure 1. NDVI image.

**Layer H2O**. This layer specifically considers the characteristic that mice do not live far from water sources; for it a supervised classification on the NDVI image was made, selecting representative areas of water bodies and wet surfaces such as rivers and lakes and its surroundings, as training areas for implementation in the algorithm Envi ©. This procedure yielded a segmented image that enhances the presence and / or absence of water.

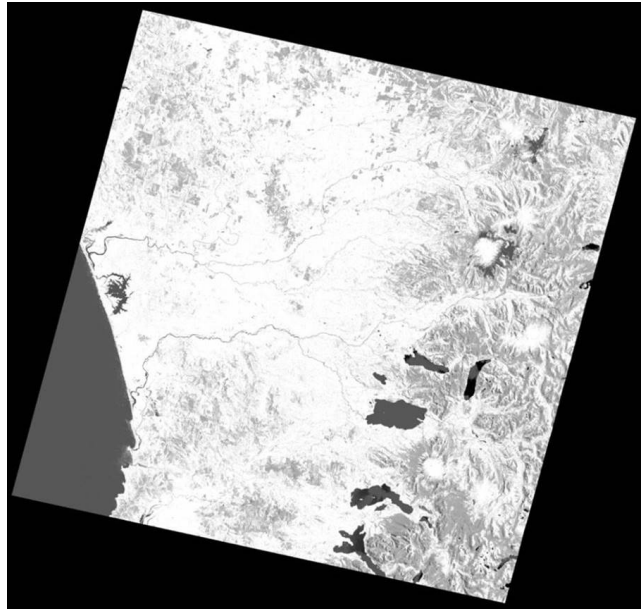

Figure 2. Moisture image derived from NDVI.

**Layer HIGH**. This information was derived from NASA GTOPO30 model. Previously it was re-dimensioned so that the number of rows and columns model completely cover the HANTA layer. Subsequently, using programming in Visual Basic © heights became likely, seeing to it that they have not found mice over 2000 meters above sea level, so the probability is set to zero percent, on the contrary heights close to the average sea level assigned one hundred percent probability.

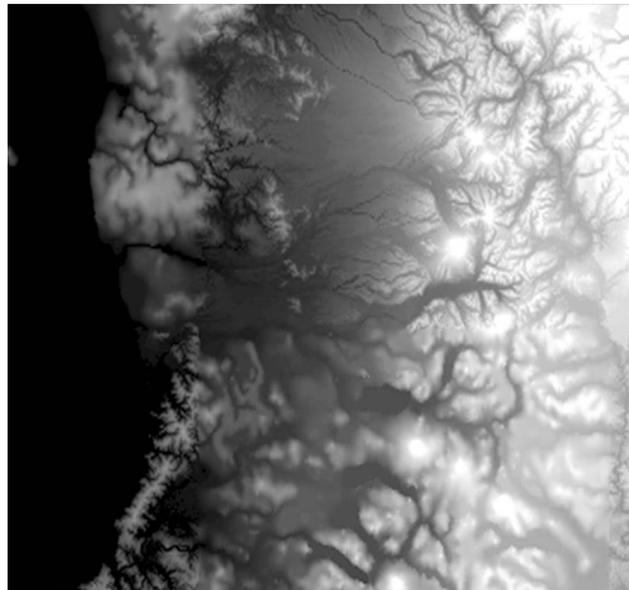

Figure 3. Digital terrain model.

**Layer MICE**. This layer is formed using the mice census information provided by the MINSAL. First, the places with seropositive mice were referenced and with this information a supervised classification was performed, so as to give greater probability of infection in areas where infected mice were actually found. In Figure 4 one can see a count derived supervised classification and cadastre of mice MINSAL image.

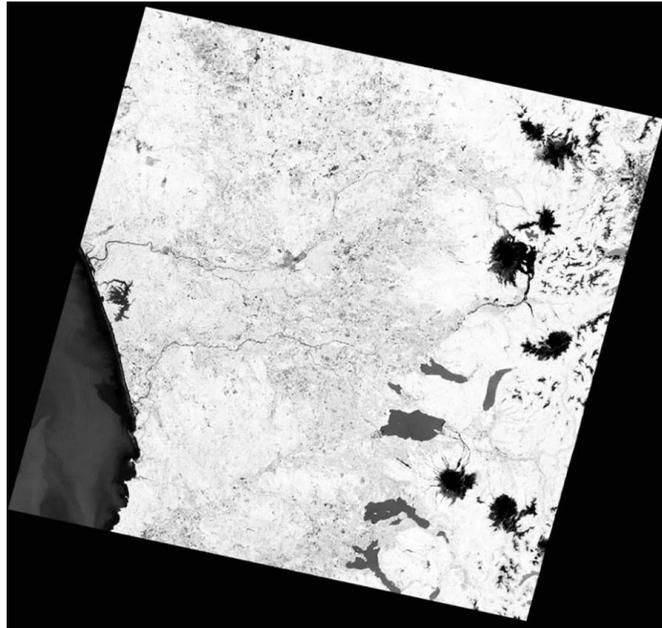

Figure 4. Supervised classification for cadastral mice.

## 5.2 Logistic regression model

Considering the information obtained, we proceeded to the creation of a first model to obtain a probability map of danger. To do so, logistic regression, which is one of the most commonly used statistical procedures for analyzing binary data, and to find the best fit with minimal parameters (*18*) was used. To calculate the parameters of the logistic regression model we used Statgraphics © software on a base of 46,274 data which correspond to 23,154 negative cases of Hanta virus and 23,120 data to positive cases, as the dependent variable HANTA layer.

Table. Estimated regression model (Maximum Likelihood)

| Parameter | Estimators | Standard Error | Probability ratio |
|-----------|-----------|----------------|-------------------|
| CONSTANT | -1,8827 | 0,0741 | |
| H2O | -0,2732 | 0,0589 | 0,7608 |
| HIGH | 2,2468 | 0,0707 | 9,4582 |
| MICE | -0,0207 | 0,0614 | 0,9794 |
| NDVI | 0,6170 | 0,0527 | 1,8533 |

Therefore, the logistic function is defined by:

$$Hanta = \frac{e^{eta}}{(1 + e^{eta})}$$

The linear function is equal to:

$$eta = -1,8827 - 0,2732 * H2O + 2,2468 * HIGH - 0,0207 * MICE + 0,6170 * NDVI$$

In Figure 5 you can see the risk map obtained after applying the logistic regression model to the data set.

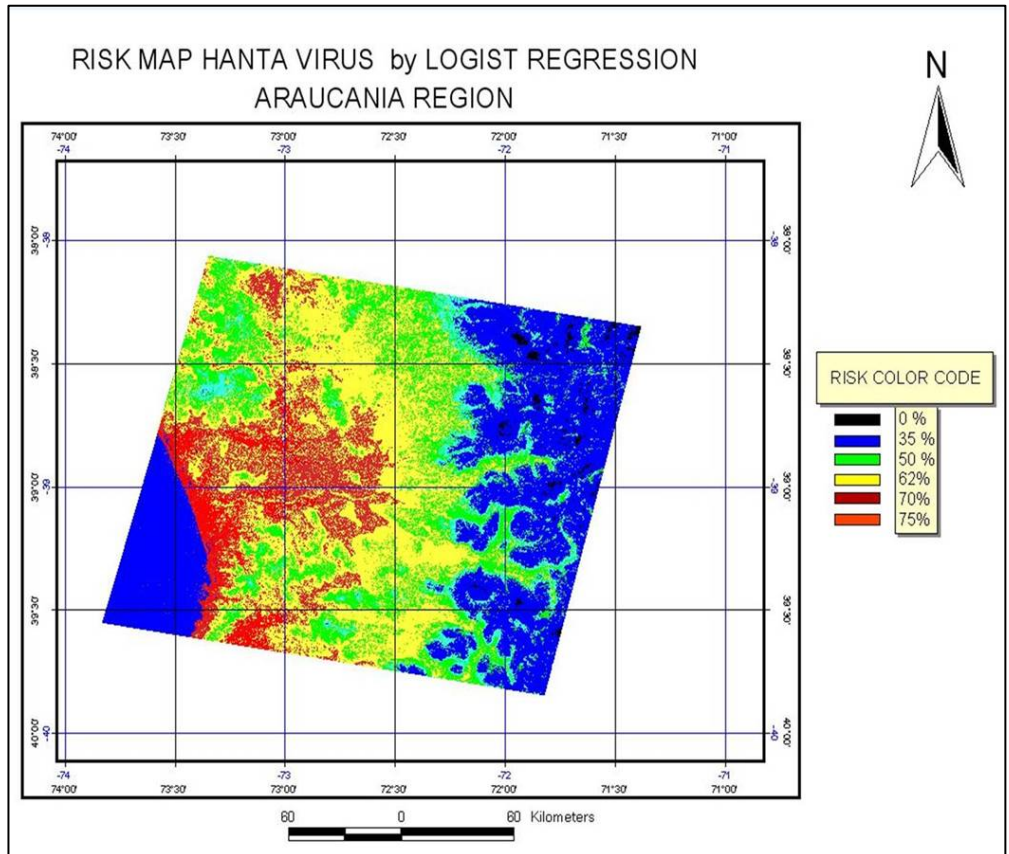

Figure 5. Risk Map Logistic Regression.

### 5.3  Artificial Neural Network (ANN)

A second map of risk probability is obtained by neural networks (ANN), which are a set of mathematical algorithms that learn from their own experience by changing their connections weights (19), commonly used in forecasting systems.

The implementation of this process was conducted with Matlab ©, whose system has implemented a series of functions to be used as decision functions, which are classified into two groups which are linear functions and non-linear functions, the latter being the used to connect more interlayers with the output neurons, thus achieving a susceptible differentiable function being minimized. Regarding the training functions, it was used back propagation, to achieve a minimum (Matlab ® toolbox).

Figure 6 shows a diagram of the final disposition of the layers of information. This information has been corrected and arranged in mathematical order in geographical coordinates.

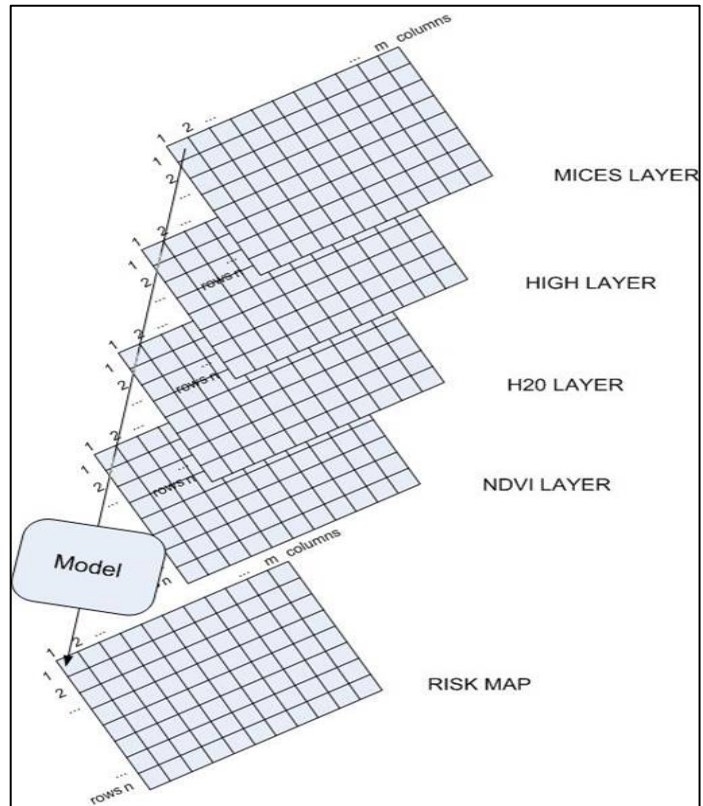

Figure 6. Data model.

For the network, we have used a first layer with 12 neurons, a second layer of 10 neurons, a third layer with 8 neurons, then 6, 4 and finally a single neuron that represents the response to training.

Using the vector entries with 240 entries including 120 positive cases and 120 negative cases proceeded to training according to the following schedule (script) made for Matlab:

```
 net = newff(minmax(N1),[12,10,8,6,4,1],{'tansig',
'tansig','tansig','tansig','tansig'},'traingdm');
net.trainParam.show =        200;
net.trainParam.Ir =          0.09;
net.trainParam.Ir_inc =      1.09;
net.trainParam.epochs =      30000;
net.trainParam.goal = 1e-3;
[net, tr] = train(net,N1,Z);
```

Where *net.trainParam.show* indicates the number of iterations for which it will present a report on training progress screen. The *net.trainParam.lr* represents the rate of learning, values between 0 and 1. The *net.trainParam.epochs* indicates how many times is presented to the ANN matrix input at this stage of training. The *net.trainParam.goal* statement indicates the desired error (the difference between the output and the desired value). The N1 matrix is the input matrix, the matrix Z contains as elements the desired output neurons values. Figure 7 shows the result of training the ANN.

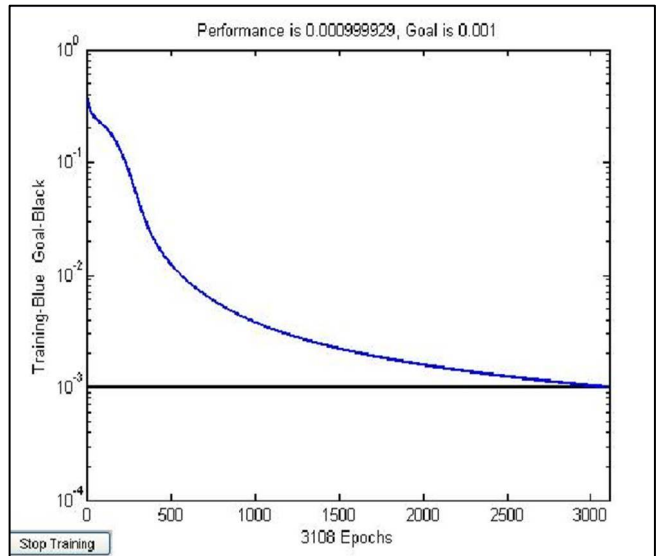

Figure 7. Training result by Backpropagation.

The construction of the arrangements was an empirical approach in which the computational processes were optimized, due to the size of the array and software limitations that arise for handling arrays with over 1500 rows x 1500 columns. The result of the process can be seen in Figure 8.

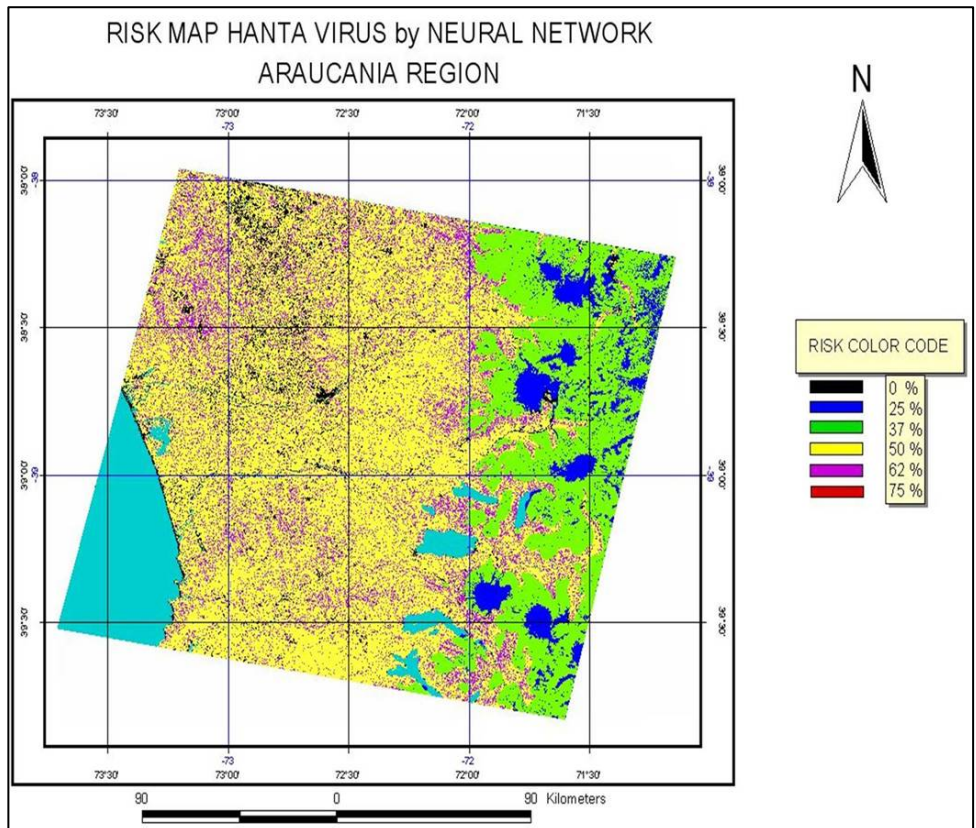

Figure 8. Risk Map using ANN.

**6. Discussion and Conclusions**

This study proposes a different look with new novel computer applications that are achievable for an application like ours with budget restrictions. Today it is possible to access Landsat images and climatic data through Internet organizations and universities, which have available scientific papers regarding geospatial information free.

In general, and performing an analysis of the results it can be seen matches in both images. In the image obtained by logistic regression greater probabilities can be seen in places with lower altitude which is logical considering that the contagion vector corresponds to the Ologorizomis longicaudatus rodent whose habitat is suitable under the 2000 MASL. This feature can also be seen in the risk map obtained by ANN. In particular, one can see characteristic features in both maps, for example in ANN classification  can be seen more homogeneous areas with very specific points with increased risk, however, in images obtained by logistic regression can be seen more diffuse risk areas but with greater definition of details in high contagion as areas near the Conguillo Park (Figure 9).

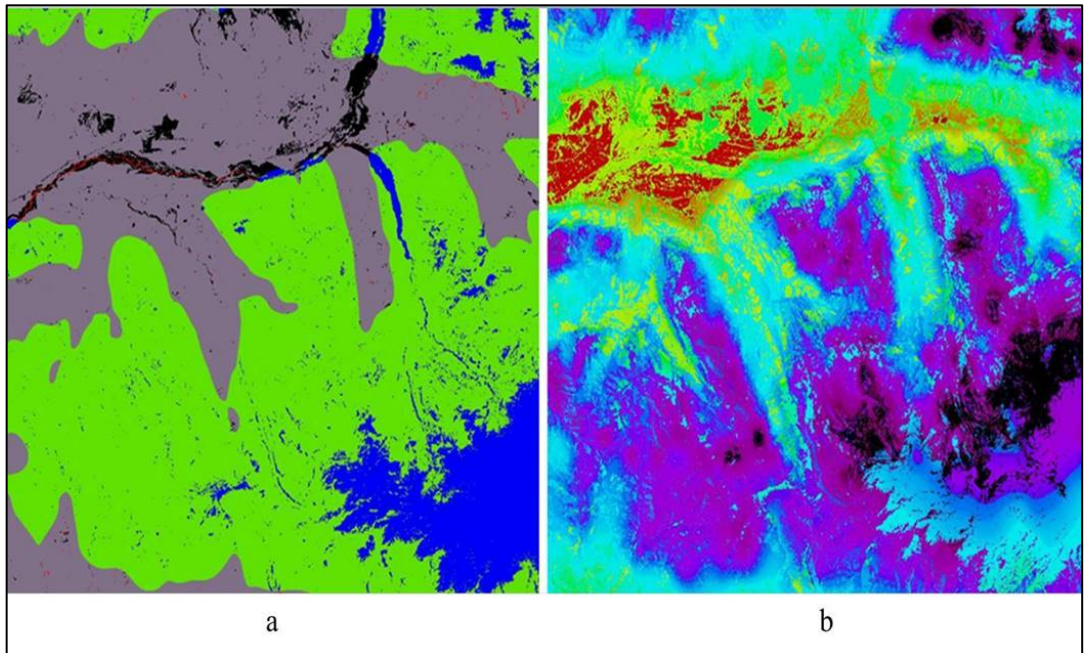

Figure 9. Detail Characterizing Risk in Sector Adjacent to Conguillío Park. a) Risk by ANN, b) Risk  by Logistic Regression.

Although the origin of the vectors for simulation belong to the same data set  (positive and negative cases of Hanta Virus) the results of both models differ in the definition of areas, appreciating better segmentation in the use of ANN. According to the authors, the nature of the phenomenon under study is better defined in the risk map image obtained by ANN. In addition, one can highlight that there is visual agreement between the two models. Importantly, the phenomenon under study and risk probability maps generated are strictly related to the origin of the data.

The main problem faced is that the data is imprecise to determine the place of infection, by the nature of the evolution of the disease; infected people usually ignore their condition until the effects are noticeable. These can appear from 48 to 72 hours after exposure. It is also important to note that this study aims to show a novel analysis method of transmission of a virus through a methodology in which automatic control tools are used with geospatial studies. In relation to the analysis of vectors, tests were performed to improve the training of ANN. Thus the matrix of positive and negative cases, which is the source to generate the model, underwent training. To do this it was necessary to divide the image into 16 subsets to process it on the ANN module.

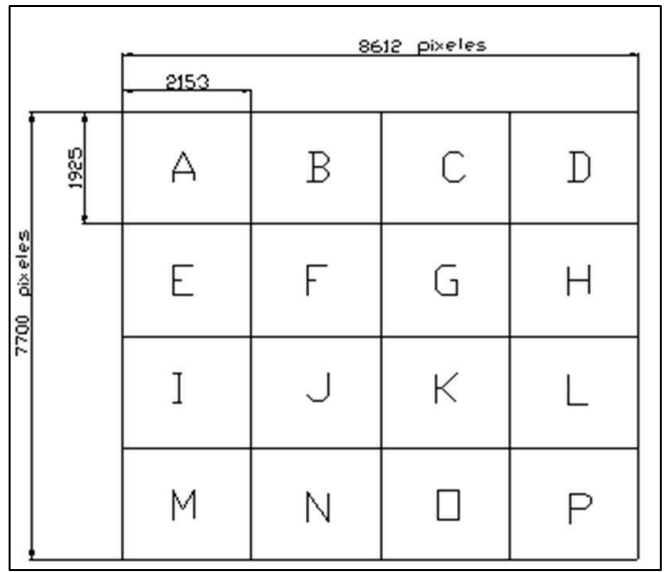

Figure 10. Subsets for data analysis.

Each subset has dimensions of 1925 rows x 2153 columns. In Figure 11 you can see the set of vectors representing the four images corresponding to the regressor variables needed to construct the model.

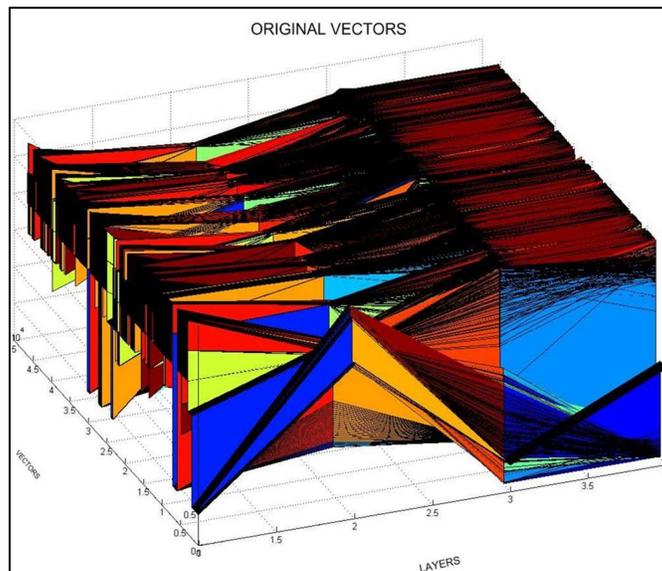

Figure 11. Original vectors.

The first training did not obtain convergence after 4000 epochs.

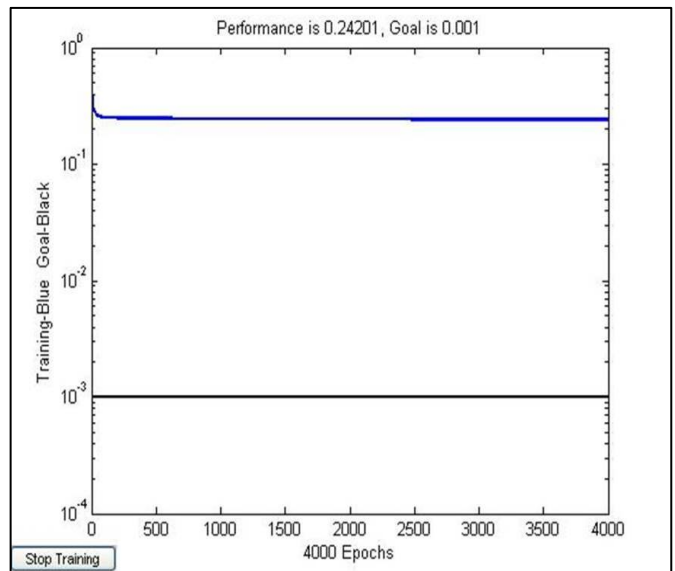

Figure 12. Non-Convergence using original vectors.

In order to improve the learning curve it was necessary to apply algorithms to adapt the segmentation of vectors and thus accelerate convergence in the training process. It was applied to the set of vectors some algorithms to highlight feature vectors in the process of generating the model with improved convergence. The first test was performed by applying a Frobenius normalization to standardize and avoid the noise produced by decimals in the process of building the model, then it was applied to the set of vectors the algorithm of fast Fourier transform. The result of this implementation can be seen in Figure 13.

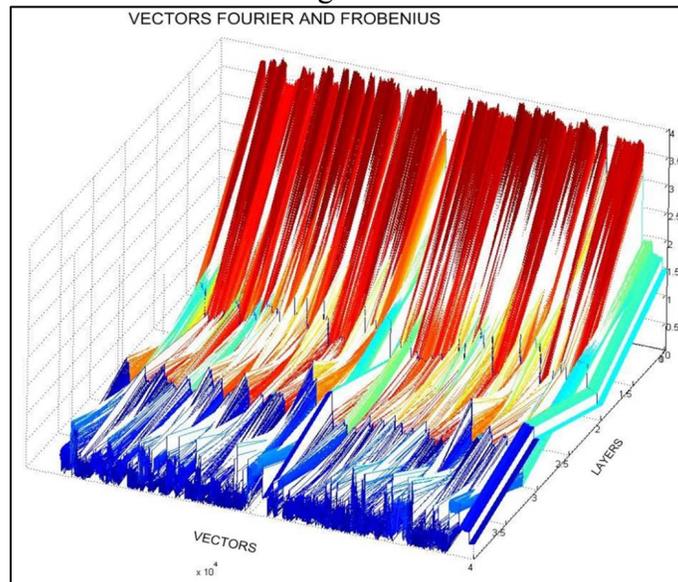

Figure 13. Application of Frobenius and Fourier to the Vector set.

The training result is shown in Figure 14.

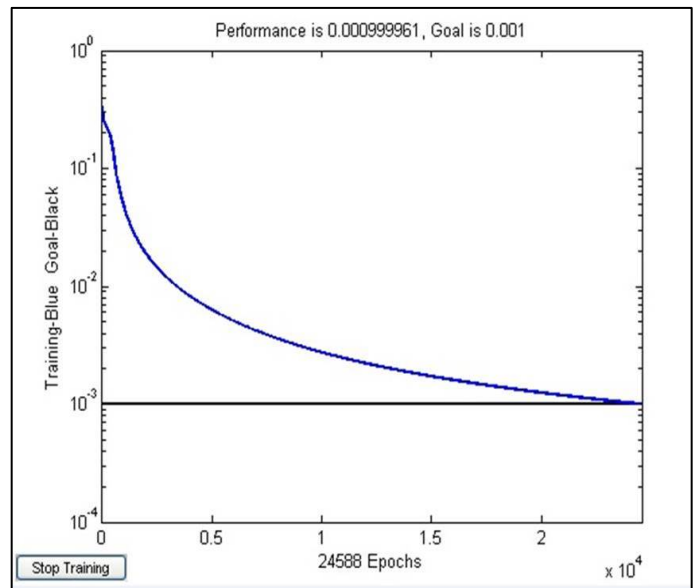

Figure 14. Convergence achieved with modified vectors.

Overall, this number of 24588 epochs favours the occurrence of overfitting (*20*).

The use of automatic control tools in environmental problems, opens up a range of possibilities in the study of the variables, with proven management vectors in an area, which at first seems distant but finally once made the necessary conversions the treatment of data is similar. Initially, the conversion of vector and matrix variables represent additional work, however, this effort increases the chances of finding models that have performed well in explaining phenomena in terms of the variables.

As a final comment, we can say that this kind of analysis requires a process of extensive sampling and a multidisciplinary team of specialists. Whereas the generation of hazard maps is not an issue that is easily communicated and probably it is not very popular with the people that lives on tourism, however, the value of the methodology still remains.

**Bibliography**
1. Rodríguez-Morales AJ. Ecoepidemiología y epidemiología satelital: nuevas herramientas en el manejo de problemas en salud pública. Revista Peruana de Medicina Experimental y Salud Pública. 2005; 22: 54-63.
2. Koch DE, Mohler RL, Goodin DG. Stratifying land use/land cover for spatial analysis of disease ecology and risk: an example using object-based classification techniques. Geospatial health. 2007; 2: 15-28.
3. Ostfeld RS, Glass GE, Keesing F. Spatial epidemiology: an emerging (or re-emerging) discipline. Trends in Ecology & Evolution. 2005; 20: 328-336.
4. Boone JD, Otteson EW, McGwire KC, Villard P, Rowe JE, St Jeor SC. Ecology and demographics of hanta virus infections in rodent populations in the Walker River Basin of Nevada and California. The American Journal of Tropical Medicine and Hygiene.1998; 59: 445-451.

5. Suzán G, Giermakowski JT, Marcé E, Suzan-Azpiri H, Armién B, Yates TL. Modeling hantavirus reservoir species dominance in high seroprevalence areas on the Azuero Peninsula of Panama. The American journal of tropical medicine and hygiene. 2006; 74: 1103-1110.

6. Ford TE, Colwell RR, Rose JB, Morse SS, Rogers DJ, Yates TL. Using satellite images of environmental changes to predict infectious disease outbreaks. Emerging Infectious Diseases. 2009; 15: 1341-1346.

7. Loehman RA, Elias J, Douglass RJ, Kuenzi AJ, Mills JN, Wagoner K. Prediction of Peromyscus maniculatus (deer mouse) population dynamics in Montana, USA, using satellite-driven vegetation productivity and weather data. Journal of wildlife diseases. 2012; 48: 348-360.

8. Andreo V, Neteler M, Rocchini D, Provensal C, Levis S, Porcasi X, et al. 2014. Estimating Hantavirus risk in southern Argentina: a GIS-based approach combining human cases and host distribution. Viruses. 2014; 6: 201-222.

9. Fan H, Ge L, Song L, Zhao Q. Analysis of the Spatiotemporal Characteristics of Hemorrhagic Fever with Renal Syndrome in Hubei Province, China. ISPRS Annals of the Photogrammetry, Remote Sensing and Spatial Information Sciences. 2015; 2: 207-210.

10. Lee HW, Lee PW, Johnson KM. Isolation of the etiologic agent of Korean hemorrhagic fever. Journal of Infectious Diseases. 1978. 137: 298-308.

11. Glass GE, Cheek JE, Patz JA, Shields TM, Doyle TJ, Thoroughman DA, et al. Using Remotely Sensed Data to Identify Areas at Risk for Hantavirus Pulmonary Syndrome. Emerging Infectious Diseases. 2000; 6: 238-247.

12. Pavletic C. Hantavirus: Su distribución geográfica entre los roedores silvestres de Chile. Revista Chilena de Infectología. 2008; 17: 186-196.

13. Sotomayor V, Aguilera X. Epidemiología de la infección humana por hantavirus en Chile. Revista Chilena de Infectología. 2000; 17:220-232.

14. Ortiz JC, Venegas W, Sandoval JA, Chandia P, Torres-Pérez F. Hantavirus en roedores de la Octava Región de Chile. Revista Chilena de Historia Natural. 2004; 77: 251-256.

15. Torres-Pérez F, Boric-Bargetto D, Palma RE. Hantavirus en Chile: Nuevos roedores con potencial importancia epidemiológica. Revista Médica de Chile. 2016; 144: 816-820.

16. MINSAL, Ministerio de Salud, Chile. http://web.minsal.cl

17. González LD, Melgar DC, Vega VS. Enfoque bayesiano del modelo de regresión logística usando cadenas de Markov Monte Carlo. Investigación Operacional. 2015; 36: 178-186.

18. Parker JR. Algorithms for Image Processing and Computer Vision. 2nd ed. Indianapolis: John Wiley & Sons, Inc.; 2011. p 364.

19. Álvarez GD, Aguilar J, Cáceres M, Salinas RA. 2001. Reconocimiento de Electroencefalogramas Epilépticos utilizando Redes Neuronales Artificiales y Análisis de FFT. XIV Congreso Chileno de Ingeniería Eléctrica Antofagasta, 19-23 noviembre. Chile. 2001.